\documentclass[12pt,preprint]{aastex}
\usepackage{psfig}
\usepackage{timesfonts}

\def\epm{\hbox{e}^\pm}

\begin{document}

\title{Iron K$\alpha$ Emission from X-ray Reflection: Predictions for
Gamma-Ray Burst Models}

\author{David R. Ballantyne and Enrico Ramirez-Ruiz}
\affil{Institute of Astronomy, University of Cambridge,
Madingley Road, Cambridge CB3 0HA, England}

\begin{abstract}
Recent observations of several $\gamma$-ray burst (GRB)
afterglows have shown evidence for a large amount of X-ray line emitting
material, possibly arising from ionized iron.  
A significant detection of an X-ray
spectral feature, such as that found in the {\it Chandra} observation
of GRB~991216, may provide important constraints on the immediate
environment of the burst and hence on progenitor models. The large
Fe~K$\alpha$ equivalent widths inferred from the X-ray observations
favor models in which the line is produced when the primary X-ray
emission from the source strikes Thomson-thick material and Compton
scatters into our line of sight. We present such reflection spectra
here, computed in a fully self-consistent manner, and discuss the
range of ionization parameters that may be relevant to different
models of GRBs. We argue that the presence of a strong hydrogen-like
K$\alpha$ line is unlikely, because Fe~{\sc xxvi} photons would
be trapped resonantly and removed from the line core by Compton
scattering. In contrast, a strong narrow emission line from He-like
Fe~{\sc xxv} is prominent in the model spectra. We briefly discuss how
these constraints may affect the line energy determination in
GRB~991216.
\end{abstract}

\section{Introduction}
The detection of spectral signatures associated with the environment
of a $\gamma$-ray burst (GRB) would provide important clues about the
triggering mechanism and the progenitor (M\'esz\'aros \& Rees 1998;
Lazzati, Campana \& Ghisellini 1999; B\"ottcher 2000). Observations
with {\it Chandra}, {\it ASCA} and {\it Beppo}SAX have provided
tentative evidence for Fe K$\alpha$ line and edge features in at
least five bursts; GRB~970508 (Piro et al. 1999), GRB~970828 (Yoshida
et al. 1999), GRB~991216 (Piro et al. 2000) and GRB~000214
(Antonelli et al. 2000) all show an emission feature during the X-ray
afterglow a few hours to a day after the burst event; while
GRB~990705 (Amati et al. 2000) displays a prominent X-ray absorption
feature during the burst itself. Although most of the line detections
are only marginally significant and fail to distinguish between the
various line excitation mechanisms\footnote{The energy of the Fe
emission line shifts up in energy from 6.4 through 6.7 to 6.97 keV as
Fe is ionized, but Doppler shifts may confuse precise estimates of the
observed energy.}, GRB~991216 shows a 3.49$\pm$0.06~keV line at a
moderate confidence level ($\sim 4 \sigma$). This is consistent with emission
from H-like Fe (Fe~{\sc xxvi}) at the redshift of the most distant
absorption system along the line of sight at $z$=1.02 (Vreeswijk et
al. 2000). A similar observation, but with higher signal-to-noise,
may be able to distinguish between the various line emission
mechanisms and lead to the correct determination of the line energy.

The large equivalent widths ($\sim$ a few keV) inferred from the X-ray
features favor models in which the line is produced by reflection
(Vietri et al. 2000; Rees \& M\'esz\'aros 2000), rather than
transmission.  Any detection of emission features in the afterglow
spectra some hours after the burst (such as in GRB~991216) therefore
imposes strong constraints on the location and geometry of the
optically thick reflecting material. Observing an X-ray line at a time
$t_{\rm obs}$ after the burst implies that the emitting material must
be located within a distance $\sim ct_{\rm obs}/(1+z)$ from the
explosion site, thus strongly limiting the size of the
remnant. However, this gas cannot be optically thick along the line of
sight because this would smear out the short-time variability of the
burst radiation. These conditions point towards a strongly anisotropic
environment from which a GRB is seen only if we happen to observe
the system through a line of sight with low optical depth (B\"ottcher
2000; Vietri et al. 2000; Lazzati et al. 2001).

Two types of reflection models have been developed to explain the
origin of the X-ray emission features. The first invokes the
interaction of the primary X-ray emission from the afterglow with an
Fe-enriched, Thomson-thick, asymmetric remnant, which Compton scatters
X-rays into our line of sight (Vietri et al. 2000; B\"ottcher \& Fryer
2000). This scenario requires a mass $\gtrsim0.06 M_{\odot}$ of Fe at
a distance about 1.5 light days, possibly due to a remnant of an
explosive event or a supernova that occurred days or weeks prior to
the GRB (see Vietri \& Stella 1998; in contrast with MacFadyen \&
Woosley 1999 which favors a simultaneous supernova explosion). The
other type of model involves a long-lived ($\gtrsim 1 {\rm \;day}$)
magnetar or accreting black hole with a continuing but decaying
outflow that interacts with the stellar envelope at distances less
than a light-hour (Rees \& M\'esz\'aros 2000; M\'esz\'aros \& Rees
2001). In this case only a small mass of Fe is required, and can be
readily produced by the star itself.

Under both interpretations, it is likely that reflection takes place
in highly ionized surfaces.  This can lead to strong Comptonization of
the emergent Fe line, and other absorption and emission features. In
this paper, we present and discuss detailed, self-consistent
computations of the temperature and ionization structure of a uniform
slab of gas ionized by the incident radiation of a GRB and of the
resulting reflection spectra. Our analysis applies to an optically
thick, homogeneous medium, significantly extending previous analysis in
the optically thin regime (e.g. Weth et al. 2000; B\"ottcher 2000),
which is hard-pressed to explain the observed Fe-line feature in
GRB 991216 (Vietri et al. 2000). We estimate the strength of the
Fe~K$\alpha$ line that each model produces. Finally, we 
discuss the implications of these results for current and
future X-ray observations.

\section{X-ray illuminated slabs}
We employ the reflection code developed by Ross, Weaver \& McCray
(1978) and updated by Ross \& Fabian (1993). We consider the
illumination of the first 12 Thomson depths of a infinite, uniform
slab of gas by radiation with a power-law spectrum of photon index
$\Gamma$ incident at an angle $\vartheta$ to the normal (Ross, Fabian
\& Young 1999). The incident radiation is treated analytically in a
`one-stream' approximation, while the diffuse radiation that results
from both the Compton scattering of the incident radiation and
emission from the gas itself, is treated using the
Fokker--Planck/diffusion method of Ross et al. (1978). Once thermal
and ionization equilibrium in the slab is found, the reflection
spectrum is computed.  We assume that hydrogen and helium are
completely ionized, but include the partially ionized species C~{\sc
v--vii}, O~{\sc v--ix}, Mg~{\sc ix--xiii}, Si~{\sc xi--xv}, and
Fe~{\sc xvi--xxvii}.

For a given $\Gamma$, the temperature and ionization state of the
surface of the slab is expected to depend mainly on the value of the
ionization parameter,
\begin{equation}
\xi={4\pi F \over n_H},
\end{equation}
where $F$ is the total illuminating flux (from 0.01-100~keV) and
$n_H$ is the hydrogen number density (Ross et al. 1999).

\section{Application to Gamma-Ray Bursts}
If the GRB explodes within a young supernova (SN) remnant, the
X-rays from the afterglow will illuminate Fe-rich ($\sim$ ten
times solar Fe abundance) material, leading to recombination line
emission by reflection (Vietri et al. 2000). Reasonable
values of $\xi$ expected in this scenario are
\begin{equation}
\xi_{SN} \approx 10^{6} L_{48}d_{16}^{-2}n_{H,10}^{-1},
\end{equation}
where $L$ is the source luminosity in units of erg s$^{-1}$, $n_H$ is
the hydrogen density in units of cm$^{-3}$, $d$ is the distance from
the burst explosion to the material in units of cm. We adopt the
convention $Q = 10^x\,Q_x$ for expressing the physical parameters. The
time delay of a few days observed in GRB~991216 yields an incident
angle for the radiation of $\vartheta \sim 45^{\circ}$ (Vietri et
al. 2000).

Alternatively, line emission can be
produced when a post-burst outflow, possibly magnetically dominated,
impacts on the walls of a funnel excavated in the stellar envelope (SE)
at distances less than a light-hour (Rees \& M\'esz\'aros
2000). Luminosities as high as $L \sim 10^{47}$ erg s$^{-1}$ are
expected 1 day after the burst, if they are due either to the spinning
down millisecond pulsar or  to a highly magnetized torus around a
black-hole (Rees \& M\'esz\'aros
2000). The ionization parameter in the stellar envelope case is
\begin{equation}  
\xi_{SE} \approx \beta 10^{4} L_{47}d_{13}^{-2}n_{H,17}^{-1}, 
\end{equation}
where $\beta <1$ is the ratio of ionizing to MHD luminosity (Rees \&
M\'esz\'aros 2000). The incident flux would be deflected along the
funnel walls with different incident angles before escaping the
funnel: for simplicity we assume $\vartheta \sim 45^{\circ}$. 

For these uniform-density slabs we vary $\xi$ by changing the total
illuminating flux while keeping the hydrogen number density $n_H$, the
distance from the burst $d$ and the incident angle $\vartheta$
fixed. Fig. 1 shows the results for illumination by a $\Gamma$=2
spectrum for both of the scenarios described above.  The illuminating
and reflected spectra are displayed as $E F_{E}$, where $F_{E}$ is the
spectral energy flux and $E$ is the photon energy. The models with the
highest ionization parameter ($\xi > 10^{4.5}$) are excellent
reflectors, and show almost no Fe spectral features. This is because
the surface layer is almost fully ionized, and Fe~{\sc xxvi} does not
become dominant until $\tau_T > 8$. The temperature at the surface of
the slab is $\sim$ 1.9 $\times 10^{7}$~K, the Compton temperature for
the incident spectrum. The spectrum drops at $\sim$ 50~keV because we
included a sharp energy cut-off in the incident spectrum at 100
keV\footnote{For a constant density slab and the values of $\Gamma$
considered here, extending the illuminating spectrum to higher
energies ($< 511$~keV) would have little effect on the iron ionization
and spectral features (e.g., Ross et al. 1999).}.  Compton reflection
produces a slight steeping in the reflected spectrum, which mimics a
power law with $\Gamma > 2$ in the 3-30 keV band. For example, we find
$\Gamma$=2.3 in the $L_{48}$ SN case and $\Gamma$=2.14 in the $L_{48}$
SE case.

When the illuminating flux is reduced so that $10^{3.5} < \xi <
10^{4}$, Fe K$\alpha$ emission and Fe K-shell absorption features
begin to appear. Most of the K$\alpha$ photons originate at $\tau_T
\gg 1$ and emerge as broad Comptonized lines with weak cores. The
H-like K$\alpha$ photons at 6.97~keV are generated close to the
surface, but are resonantly trapped and removed from the narrow line
core by Compton scattering. The He-like inter-combination line at
6.7~keV is not subject to resonant trapping, but is generated at such
high $\tau_T$ (see panel (b) in Fig. 2) that it is multiply Compton
scattered on leaving the slab.  The broad Comptonized emission
features blended into the Compton smeared K-shell absorption edge, are
an important signature of ionized reflection. The effect of increasing
the iron abundance at a fixed ionization parameter is illustrated by
the dotted line in Fig. 1.  The Comptonized line and the absorption
feature are increasingly significant for an Fe-rich medium.

With $\xi \sim 10^{3}$, Fe~{\sc xxv} becomes dominant at $\tau_T
\approx 1$ (see panel (a) in Fig. 2). Narrow emission line resulting
from Fe~{\sc xxv} can now be seen superimposed on the
Compton-broadened emission bump. The tiny emission feature just above
8.8 keV results from radiative recombination directly to the ground
level of Fe~{\sc xxv} (see the $L_{46}$ SE case in Fig 2). In the
model with $\xi_{SE}=10^{2}$ the Fe emission is suppressed because
Fe~{\sc xvii-xxii} dominates close to the surface, but their K$\alpha$
photons are destroyed by the Auger effect during resonance trapping
(Ross, Fabian \& Brandt 1996). Finally, for $\xi_{SE}=10$ the
reflection spectrum is similar to that of a cold, neutral slab, and so
the narrow emission line at 6.4 keV is dominant.

The effect of increasing the incident angle $\vartheta$, which is a
key ingredient in the pre-ejection models because it is responsible
for the observed time-delay, is also illustrated in Fig. 1 by the
dot-dashed line. Radiation that illuminates the atmosphere more
directly ionizes more deeply into the slab than radiation at grazing
incidence. At high flux levels the emergent line features are little
changed, however, the K-shell absorption features become more
prominent for radiation that impacts closer to the normal of
the slab.

The ionization structure in the outer layers of the illuminated slab
also depends on the incident radiation spectrum. Fig. 3 shows the Fe
K$\alpha$ equivalent width (EW) as a function of incident luminosity
for a series of reflection spectra with $\Gamma=$1.6, 2.0 \& 2.4,
assuming $\vartheta=45^{\circ}$. The EWs were calculated with respect
to the reflection spectrum, and the integration was carried out
between 5.7 and 7.1~keV. As seen in Fig. 1, the EW of the Fe K$\alpha$
line decreases with $L$ in both GRB scenarios. This behavior continues
until $L$ is sufficiently low for the narrow Fe K$\alpha$ emission
line to be suppressed by the Auger effect. The EWs are generally
larger when the illuminating spectrum is steeper (i.e., $\Gamma$ is
greater). The weaker ionizing power of steep spectra allow line
emission to persist at the highest luminosities, although the EWs
still end up quite low when $L=L_{48}$. In a GRB, softer spectra may
be important if both the line emission arises from the impact of the
continuous energy output with the compact remnant and much of the
observed continuum emission (with a flatter $\Gamma$) comes from the
afterglow emission directly.

\section{Discussion}

The emission feature observed $\sim$ 1.5 days after the GRB~991216
burst (Piro et al. 2000) had a line luminosity of $L_{\rm line} =4
\times 10^{44}$ erg s$^{-1}$ and an equivalent width (EW) of $\sim$
0.5 keV. The continuum flux from GRB~991216, measured to have
$\Gamma=2.2 \pm 0.2$, in the 1--10 keV band was 50--100 times stronger
than the flux in the line.  As is clear from the above discussion, the
emission feature can be explained by reflection if identified with the
recombination K$\alpha$ line from He-like iron at 6.7 keV.
 
Taking $\Gamma \approx 2$ in the SE scenario, a continuous ionizing
luminosity of $\sim 10^{45}-10^{46}$ erg s$^{-1}$ (or a continuous
wind luminosity of $10^{47}$ erg s$^{-1}$ with $\beta \sim 0.01-0.1$;
see Rees \& M\'esz\'aros 2000) would be sufficient to produce the
observed line\footnote{Under this interpretation, the observed line
width can easily be reproduced by Comptonization for an expansion
velocity below the limit of 0.1$c$ inferred by Piro et
al. (2000).}. However, it is possible that the reflected and incident
spectra are observed together. With the above parameters, the line
luminosity and EW calculated for the total spectrum (reflected +
incident) are $L_{\rm line} \sim 2 \times 10^{44}$ erg s$^{-1}$ and EW
$\sim$ 0.25--0.7 keV.  If the illuminating spectrum is a steeper
power-law ($\Gamma >$ 2), then a smaller fraction of illuminating
photons lie in the 9--20 keV range which dominates the ionization of
Fe, and thus an increase in either $L$, $\beta$ or the Fe abundance is
required in order to reproduce the same line strength (Fig. 3).

Alternatively, an X-ray afterglow with $\Gamma \approx 2$ illuminating
the walls of a supernova remnant with a luminosity of $\sim
10^{45.5}-10^{46}$ would produce a line signal with $L_{\rm line}
\sim 2 \times 10^{44}$ erg s$^{-1}$ and EW $\sim$ 0.2--0.8
keV. However, an essential assumption contained in this model is that
the material responsible for emission-line features is
illuminated by the early afterglow radiation (or the GRB itself) with
$L \ge L_{48}$. Such high luminosities would cause the spectral
features resulting from iron to disappear or, at best, to be extremely
weak (even for larger values of $\Gamma$, see Fig. 3). Moreover, this
early incident radiation can  be harder than a $\Gamma \sim 2$
spectrum with a significant fraction of the energy above the
$\gamma\gamma \to \epm$ formation energy threshold, and a
high compactness parameter. This will cause new pairs to be
formed in the originally optically thick scattering medium, an effect
which amplifies the density of scattering charges and increases the
temperature of the illuminating material. When pairs are produced in
sufficient numbers, the iron K$\alpha$ emission is suppressed due to 
the decrease in the number of recombinations. These effects will be
investigated elsewhere (Lazzati, Ramirez-Ruiz \& Rees 2001, in
preparation). These problems may be overcome in
particular source geometries for which lower luminosities and softer
spectra are expected at the edges of the relativistic outflow.

The results presented in Fig.~1 show that if the observed Fe K$\alpha$
line from GRB~991216 is identified solely with the H-like line at 6.97
keV, as suggested by Piro et al. (2000), then this is inconsistent
with emission from a photoionized optically thick
slab. Their detailed analysis of the {\it Chandra} ACIS-S spectrum
shows marginal evidence (at the $2.1\sigma$ level) of an emission feature at
4.4 $\pm$ 0.5 keV. Identifying this feature with the Fe recombination
edge with rest energy of 9.28 keV gives $z$= 1.1 $\pm$ 0.1, consistent
with the redshift of a H-like Fe K$\alpha$ line. Nonetheless,
it is important to emphasize that the observations presented by Piro
et al. (2000) do not rule out the presence of a He-like Fe feature
(or a blend between H and He-like features as shown in Fig. 1).
Some ambiguity also remains in the rest energy of the emission since
Doppler blue-shifts of order $0.1c$ are expected in both reflection
scenarios (Vietri et al. 2000; Rees \& M\'esz\'aros 200; M\'esz\'aros
\& Rees 2001) and  may confuse precise measurements of the line
energy. Indeed, for an expansion velocity of $0.1c$, the
emission feature at 4.4 $\pm$ 0.5 keV could be attributed to the
Fe~{\sc xxv} radiative recombination emission just above 8.8 keV
rather than to the Fe recombination edge at 9.28 keV.  

Finally, Co produces both a He and H-like emission line: at
7.242 keV and just above 7.526 keV, respectively. The exact strength of
these features, which we do not include in our calculations, could be
important in both of the scenarios discussed here and may confuse the
identification of low signal-to-noise spectral features. 

We have presented calculations of X-ray reflection from Thomson-thick
slabs for conditions which may arise in the immediate environment of a
GRB. Comparisons between putative Fe K$\alpha$ lines detected in
X-ray afterglows and those predicted by the computations will be
useful in distinguishing between the various line emission
mechanisms. We expect that more sensitive  data on X-ray afterglow
spectral features will impose strong constraints on the nature of GRB
progenitors and their environments.

\begin{acknowledgements}
We thank M. J. Rees, P. M\'esz\'aros, D. Lazzati, A. Blain and the
referee for useful comments and suggestions. We are particularly grateful to
A. C. Fabian and R. R. Ross for very helpful insights regarding the
calculations.  ERR acknowledges support from CONACYT, SEP and the ORS
foundation.  DRB thanks the Commonwealth Scholarship and Fellowship
Plan and the Natural Sciences and Engineering Research Council of Canada for
support. 
\end{acknowledgements}

\clearpage

\centerline{\psfig{file=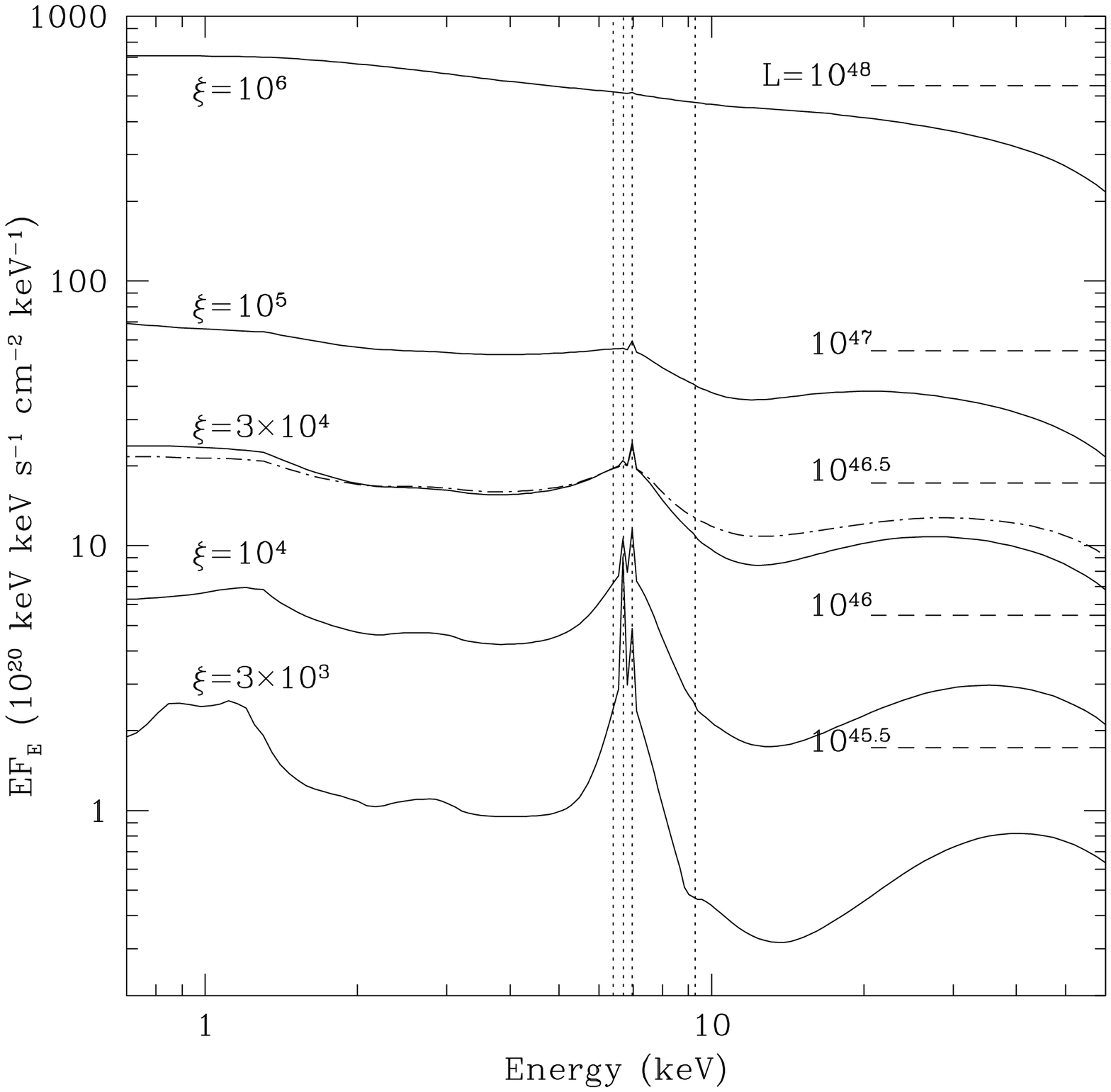,width=0.45
\textwidth}\psfig{file=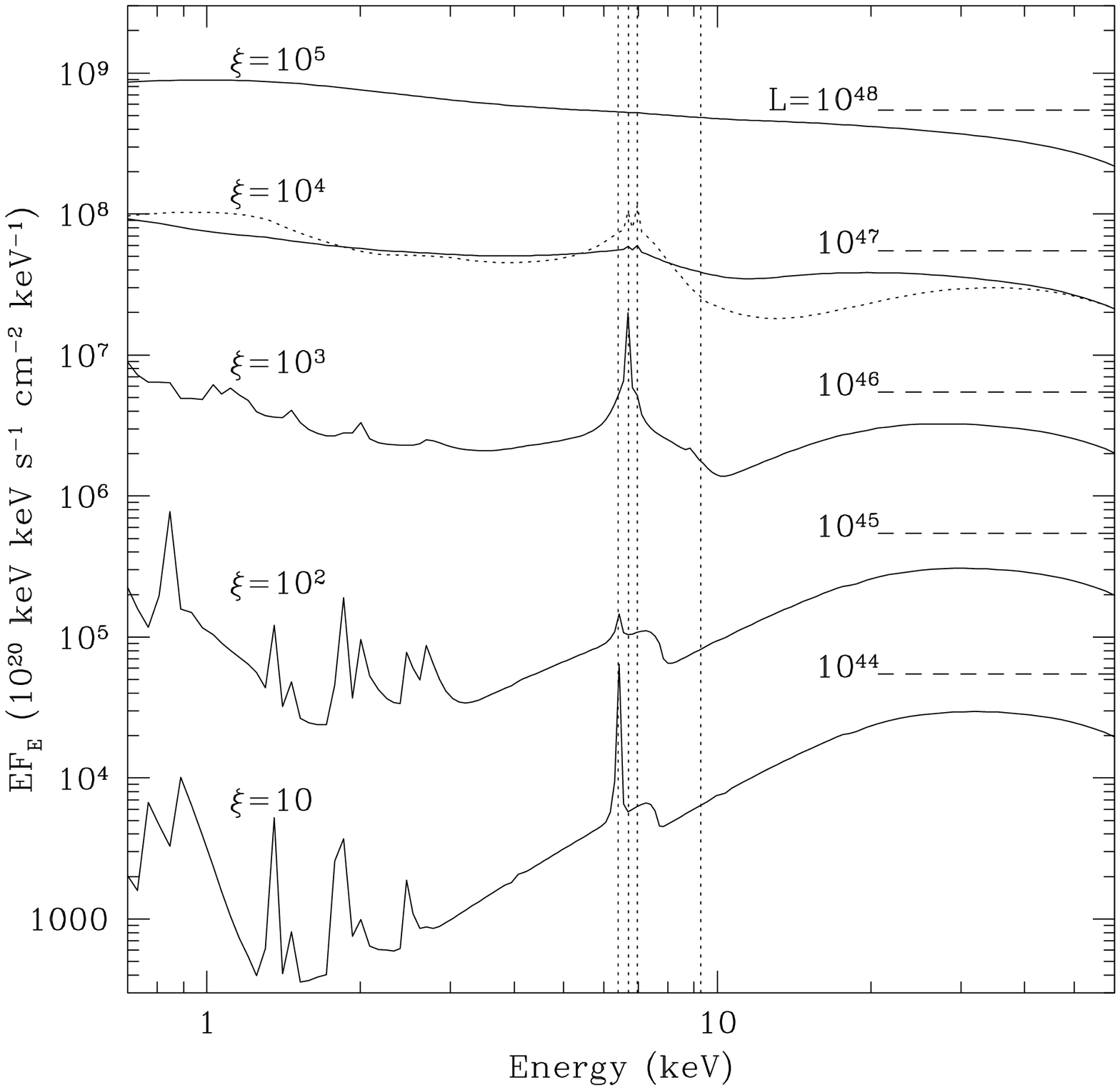,width=0.45 \textwidth}}
\figcaption{{X-ray reflection spectra for illumination with a
$\Gamma$=2 power law. In both panels the vertical lines are at (from left to right) 6.4, 6.7 and 6.97~keV for the three different Fe K$\alpha$ line energies, and 9.28~keV which is the energy of the iron recombination edge. \textbf{Left panel:} illumination of a uniform supernova
remnant with ten times solar Fe abundance, various values of the
afterglow luminosity, and $\vartheta=45^{\circ}$. The ionization
parameter is given by $\xi_{SN}= {10^{6}} L_{48}$, assuming
$d_{16}$=1, $n_{H,10}$=1 (see Eq. 2).  From top to bottom the
equivalent widths and the luminosities (integrated between 5.7 and
7.1~keV) inferred from the X-ray line features are: 3~eV, 37~eV,
94~eV, 268~eV, 943~eV; 2.5$L_{46}$, 2.7$L_{45}$, 9.6$L_{44}$,
3.6$L_{44}$, 1.4$L_{44}$. The dot-dashed line illustrates the effect
of increasing the incident angle from $\vartheta=45^{\circ}$ to
$\vartheta=75^{\circ}$ at a constant luminosity. \textbf{Right
panel:} illumination of a funnel excavated in the stellar envelope with
solar Fe abundance and various values of the decaying X-ray luminosity
from the GRB central engine. In this case $\xi_{SE}= {10^{5}} L_{48}$,
assuming $d_{13}$=1, $n_{H,17}$=1, $\beta$=1 (see Eq. 3) with
$\vartheta=45^{\circ}$. From top to bottom the equivalent widths and
the luminosities inferred from the X-ray line features are: 3~eV,
62~eV, 1.1~keV, 122~eV, 1.6~keV; 2.6$L_{46}$, 2.7$L_{45}$,
2.7$L_{44}$, 5$L_{42}$, 5.2$L_{41}$. The dotted line shows the
reflected spectra for illumination with the same luminosity but into a
stellar envelope that is ten times more abundant in iron. The line
luminosity and equivalent width for this model are: 243~eV and
3.7$L_{45}$.} 
\label{fig1a}}

\clearpage

\centerline{\psfig{file=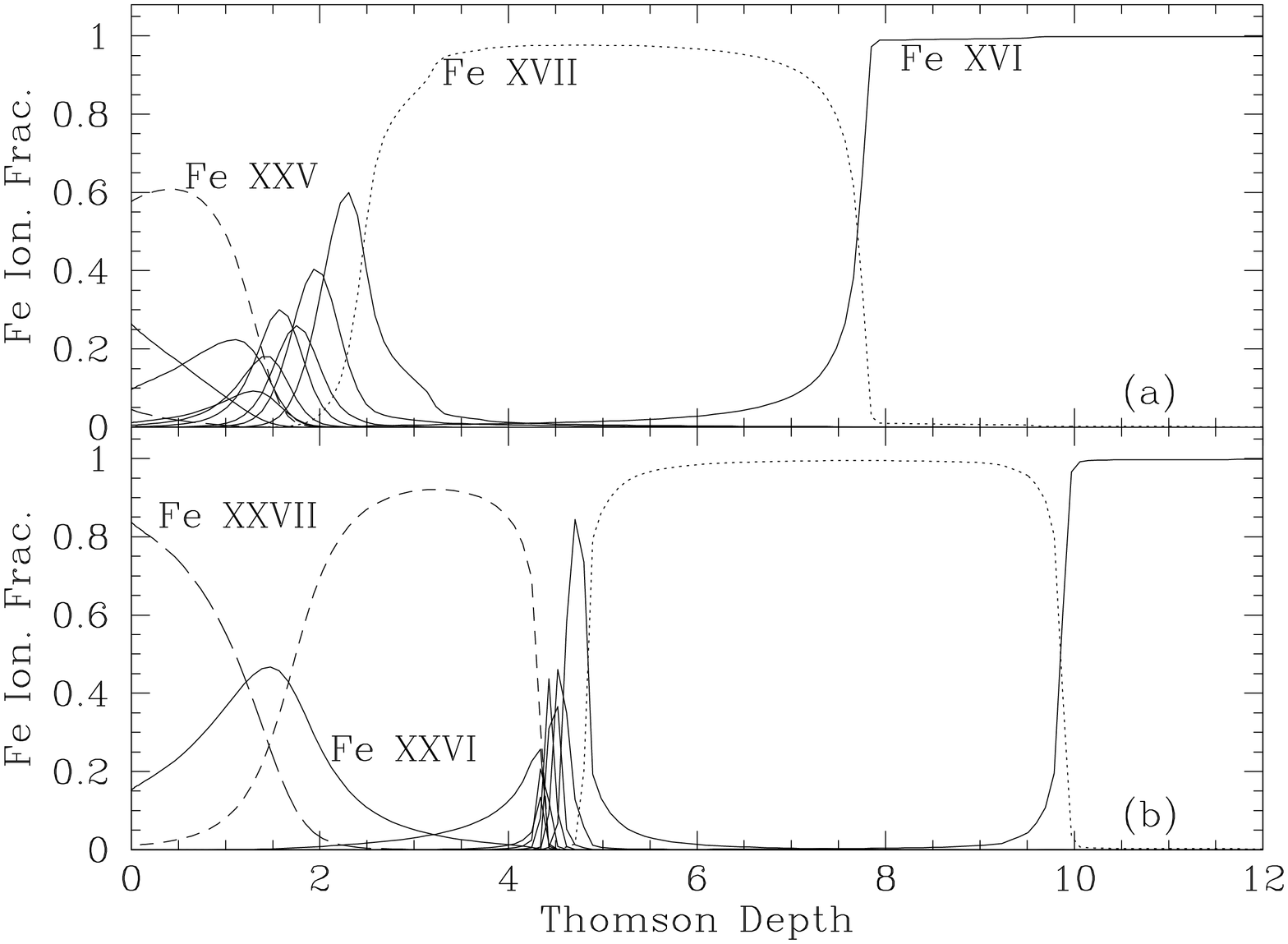,angle=-90,width=0.6 \textwidth}}
\figcaption{{Fe ion fractions as a function of Thomson depth produced
by the illumination of a uniform slab with $\Gamma$=2. The same line
type denotes the same species of Fe in both plots. Panel (a):
illumination of a funnel excavated in the stellar envelope with solar
Fe abundance and $\xi_{SE}=10^{3}$.  (b):
illumination of a uniform supernova remnant with ten times solar Fe
abundance and $\xi_{SN}={10^{4}}$.}
\label{fig2}}

\clearpage

\centerline{\psfig{file=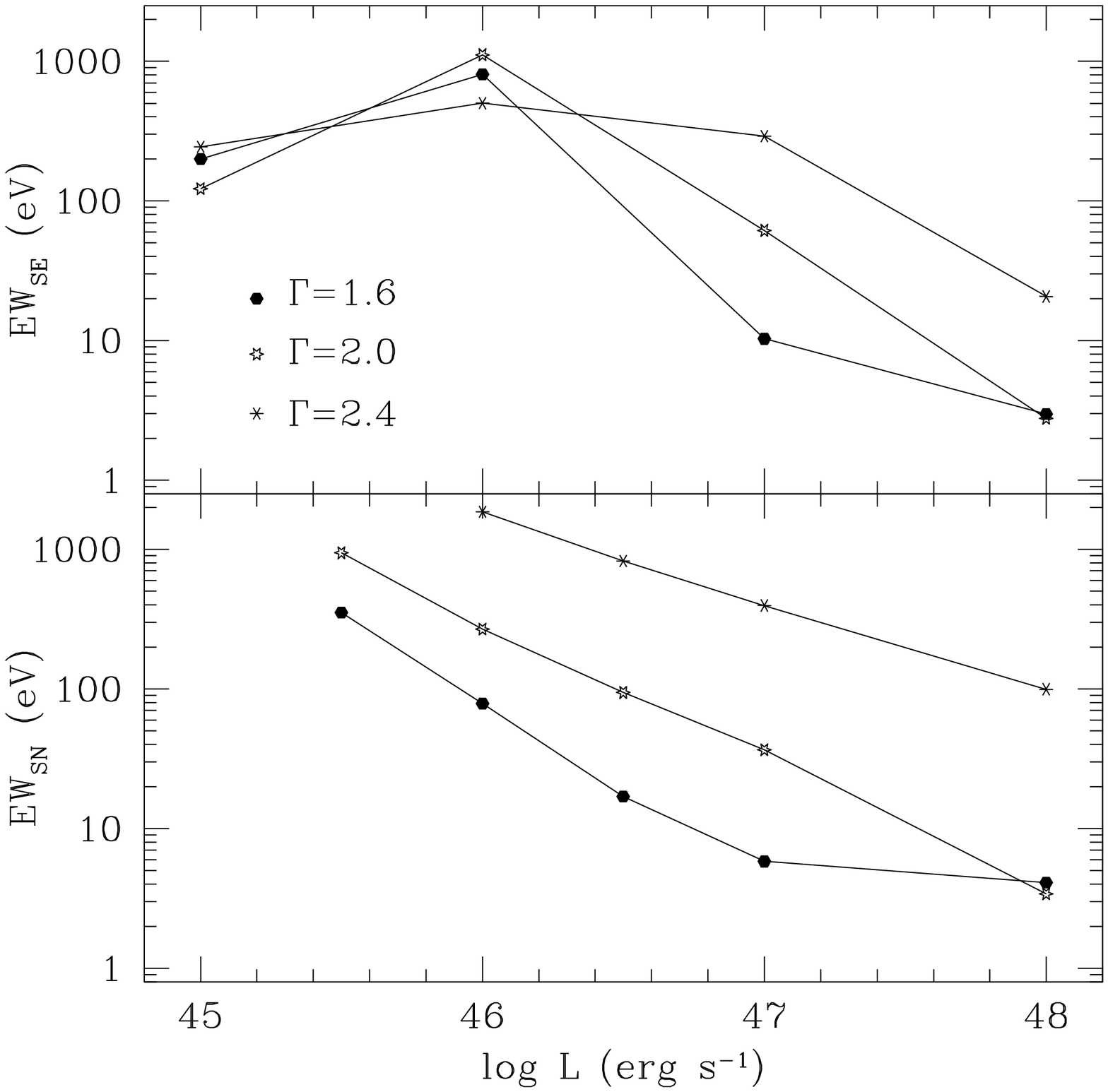,width=0.6 \textwidth}}
\figcaption{{Fe K$\alpha$ equivalent widths (EW) as a function of both
the incident luminosity and $\Gamma$ for the SE \& SN
scenarios. $\vartheta$ was taken to be 45$^{\circ}$ for these
models. The EWs were computed from the calculated reflection spectra
(integrated between 5.7 and 7.1~keV). If one uses the total spectrum
(incident + reflected) then smaller EWs are obtained. For example, the
EWs for $\Gamma$=1.6, 2.0 and 2.4 when $L=L_{46}$ (in both cases) are
370~eV, 430~eV \& 131~eV (SE) and 43~eV, 145~eV \& 805~eV (SN).}
\label{fig3}}

\end{document}